\documentclass[aps,prd,twocolumn]{revtex4}
\usepackage{amsmath,amssymb,amsfonts}
\usepackage{graphicx}
\usepackage{enumerate} 
\usepackage{colordvi} 
\usepackage{bm}
\usepackage{color}

\newcommand{\be}{\begin{eqnarray}}
\newcommand{\ee}{\end{eqnarray}}
\newcommand{\simgt}{\lower.5ex\hbox{$\; \buildrel > \over \sim \;$}}
\newcommand{\simlt}{\lower.5ex\hbox{$\; \buildrel < \over \sim \;$}}

\newcommand{\bfx}{\boldsymbol{x}}
\newcommand{\bfv}{\boldsymbol{v}}

\newcommand{\bfk}{\boldsymbol{k}}

\newcommand{\cyc}{({\rm cyc.})}

\newcommand{\Omegam}{\Omega_{\rm m}}
\newcommand{\Omegab}{\Omega_{\rm b}}

\newcommand{\rmd}{{\rm d}}

\newcommand{\hmpci}{h\,\mathrm{Mpc}^{-1}}
\newcommand{\himpc}{h^{-1}\mathrm{Mpc}}
\newcommand{\tdelta}{\tilde{\delta}}

\begin{document}
\title{Testing the equal-time angular-averaged consistency relation of the gravitational dynamics in $N$-body simulations}
\vfill
\author{Takahiro Nishimichi$^{1}$, Patrick Valageas$^{2}$}
\bigskip
\affiliation{$^1$Institut d'Astrophysique de Paris \& UPMC (UMR 7095), 98, bis boulevard Arago , 75014, Paris, France}
\affiliation{$^2$Institut de Physique Th\'eorique, CEA Saclay, 91191 Gif-sur-Yvette, cedex, France\\
CNRS, URA 2306, 91191 Gif-sur-Yvette, cedex, France}
\bigskip
\date{\today}
%

\begin{abstract}
We explicitly test the equal-time consistency relation between the angular-averaged 
bispectrum and the power spectrum of the matter density field, employing a large suite
of cosmological $N$-body simulations.
This is the lowest-order version of the relations between $(\ell+n)$-point 
and $n$-point polyspectra, where one averages over the angles of $\ell$ soft modes.
This relation depends on two wave numbers, $k'$ in the soft domain and $k$ in the hard 
domain. We show that it holds up to a good accuracy, when $k'/k\ll 1$ and 
$k'$ is in the linear regime, while the hard mode $k$ goes from linear ($0.1\,\hmpci$) 
to nonlinear ($1.0\,\hmpci$) scales.
On scales $k\lesssim 0.4\,\hmpci$, we confirm the relation within the statistical error of the simulations
(typically a few percent depending on the wave number), even though the bispectrum can already deviate 
from leading-order perturbation theory by more than $30\%$.
We further examine the relation on smaller scales with higher resolution simulations.
We find that the relation holds within the statistical error of the simulations at $z=1$,
whereas we find deviations as large as 
$\sim 7\%$ at $k \sim 1.0\,\hmpci$ at $z=0.35$. 
We show that this can be explained partly by the breakdown of the approximation
$\Omegam/f^2\simeq1$ with supplemental simulations done in the Einstein-de Sitter background cosmology.
We also estimate the impact of this approximation on the power spectrum and bispectrum.
\end{abstract}

\maketitle

\maketitle
\flushbottom
\section{Introduction}
The large-scale structure of the Universe provides us with a wealth of information
on the initial conditions of the Universe as well as the underlying gravity theory
that governs the time evolution on sufficiently large scales 
\cite{Peebles1993,Laureijs2011}.
A classic tool for discussing its statistical properties are the polyspectra of the
matter density field at a given time (the Fourier transforms of the $n$-point
correlation functions) \cite{Peebles1980,Bernardeau02}.
The power spectrum, the lowest-order polyspectrum,
has played a central role to test cosmological models and determine their
parameters precisely. Standard models of the early Universe predict 
almost Gaussian initial conditions, in agreement with a number of observational probes
(e.g., measures of cosmic microwave background anisotropies \cite{Planck_nG}). 
However, even if the initial conditions are perfectly Gaussian, the cosmic density field
at late times exhibits non-Gaussian features acquired through the nonlinear
gravitational dynamics.

The polyspectra induced by gravity can be analytically derived order by order using
standard perturbation-theory techniques (see Ref.~\cite{Bernardeau02} for a review).
In these calculations, an approximate treatment is usually adopted that greatly
simplifies the structure of the basic equations. That is, the combination 
$\Omegam/f^2$ is replaced with unity, where
$\Omegam$ is the time-dependent matter density parameter and 
$f\equiv\mathrm{d}\ln D_+/\mathrm{d}\ln a$ is the linear growth rate, with $D_+$ 
being the linear growing mode.
This approximation is exact in the Einstein-de Sitter universe and sufficiently accurate
in most other cosmological models based on general relativity, because i) one usually
recovers Einstein-de Sitter at early times and ii) over the realistic range of cosmological
parameters one has $f \simeq \Omega_{\rm m}^{\gamma}$ with $\gamma \simeq 0.5$
\cite{Peebles1980}.
When this approximation is applied, all the dependence on the cosmological 
parameters is absorbed by the linear growth rate $D_+$, and the time 
dependence of the solution is also fully encapsulated in $D_+$.
This simplifies perturbative computations because one can factor the time and
scale dependence of high-order diagrams (e.g., the contribution of order $n$ to the
power spectrum scales as $D_+^{2n}$).

Beyond perturbation theory, several articles have recently been devoted to the study
of exact ``consistency relations'' that remain valid in the nonperturbative regime,
independently of the small-scale physics (including baryon or star-formation processes)
\cite{Kehagias:2013lr,Peloso:2013fk,Creminelli:2013qy,Kehagias:2013uq,Peloso:2013fj,Creminelli:2013kx,Valageas:2013ls,Creminelli:2013yq}. 
They relate the $(\ell+n)$-point correlation, with $\ell$ modes in the linear regime 
(soft domain) and $n$ modes at much higher wave numbers (hard domain)
that can be in the nonlinear regime, to the $n$-point correlation (with $\ell$ 
linear power spectrum prefactors).
These results can be interpreted as the response of small structures 
(i.e., each element in the cosmic web such as walls, filaments or halos) to an
initial density perturbation on much larger scales.
More precisely, they derive from the equivalence principle, which ensures that
all particles and structures fall in the same fashion in a gravitational potential
force with a constant gradient. Then, at leading order, a large scale perturbation of
the initial conditions merely transports smaller scale structures without distortions.
Thus, a detection of a violation of these consistency relations would signal
a deviation from Gaussian initial conditions, significant decaying modes, or 
a departure from general relativity.
 
In the standard scenario, the kinematic consistency relations discussed above vanish
at equal times (because equal-time statistics cannot distinguish a uniform translation
of the system).
By going to the next order, and taking an angular average over the soft modes,
Refs.~\cite{Valageas:2013lr,Kehagias:2013fk} derived angular-averaged consistency
relations that remain nontrivial even at equal times.
Because this involves the dynamics of small-scale structures in a 
gravitational potential with a uniform curvature (the order beyond a constant gradient),
this probes the physics beyond the equivalence principle and it is sensitive to the
details of the dynamics. In particular, the explicit relations one obtains only hold for
dark matter (i.e., they would be violated by nongravitational processes) and within
the approximation $\Omegam/f^2 \simeq 1$, which enables us to relate the dynamics
associated with different backgrounds (which correspond to different large-scale
curvatures).
However, within these approximations they remain valid in the nonperturbative regime.

In this study, we examine the validity of these angular-averaged relations by employing
a large set of cosmological $N$-body simulations.
We focus on the lowest-order consistency relation for the angular-averaged matter 
bispectrum, which is the most interesting one in practice. 
Reference~\cite{Valageas:2013lr} has already checked this relation for the bispectrum
explicitly at the leading order of perturbation theory. 
The aim of this study is to see how higher-order corrections enter both sides of
the equation and how accurately the relation is recovered on smaller scales
(i.e., whether and by how much nonlinearities amplify the inaccuracy due to the
approximation $\Omegam/f^2 \simeq 1$).

This paper is organized as follows. We briefly review the angular-averaged consistency
relations and their validation at tree level in Sec.~\ref{sec:the_equation}. 
We then present the simulation analysis in Sec.~\ref{sec:sim}, starting form the detail
of the simulations in Sec.~\ref{subsec:nbody} and next showing our results for
the consistency relation in Sec.~\ref{subsec:results}.
We finally give a summary of the paper in Sec.~\ref{sec:summary}.
The effect of binning in the measurements of the spectra as well as a 
convergence study of the simulations are respectively presented in Appendixes~\ref{app:binning} and \ref{app:conv}.
 
\section{Angular-averaged consistency relations} 
\label{sec:the_equation}
We briefly summarize the angular-averaged consistency relations in this section.
We also review the perturbative expressions for the relevant spectra here.
\subsection{General cases} 
\label{subsec:general}
Because of statistical homogeneity, polyspectra contain a Dirac factor 
$\delta_\mathrm{D}$ that we can factor out by defining
\be
\langle \tdelta_{\bfk_1} \dots \tdelta_{\bfk_n} \rangle 
= \delta_\mathrm{D} ( \bfk_1 + \dots + \bfk_n ) \;
\langle \tdelta_{\bfk_1} \dots \tdelta_{\bfk_n} \rangle'  ,
\label{eq:polyspec}
\ee
where $\langle \dots \rangle$ is the statistical average over the Gaussian initial 
conditions and the prime in $\langle \dots \rangle'$ denotes the average in Fourier
space without the Dirac factor.
We denote the nonlinear density contrast in Fourier space by $\tdelta$, with a wave 
vector shown by the subscript. 
In a similar fashion, we also consider mixed spectra, 
$\langle \tdelta_{L,\bfk_1'} \dots \tdelta_{L,\bfk_{\ell}'} 
\tdelta_{\bfk_1} \dots \tdelta_{\bfk_n} \rangle$, which cross-correlate the nonlinear
density contrast $\tdelta$ with the linear density contrast $\tdelta_L$.
Here, $\tdelta_L$ is the linear growing mode that also defines the Gaussian
initial conditions (we assume as usual that decaying modes have had time to
become negligible).

Integrating over the direction of the linear wave numbers $\bfk_j'$, we introduce the 
angular-averaged mixed polyspectra by
\be
\int\prod_{j=1}^{\ell}\frac{\mathrm{d}\Omega_{\bfk'_j}}{4\pi}\, 
\langle \tdelta_{L,\bfk_1'} \dots \tdelta_{L,\bfk_{\ell}'} 
\tdelta_{\bfk_1} \dots \tdelta_{\bfk_n} \rangle'_{k'_j\to0} = 
\nonumber\\
\langle \overline{ \tdelta_{L,\bfk_1'} \dots \tdelta_{L,\bfk_{\ell}'} }
\tdelta_{\bfk_1} \dots \tdelta_{\bfk_n} \rangle'_{k'_j\to0}  ,
\label{eq:angleaverage}
\ee
where the limit $k'_j\to0$ is taken for all the $\ell$ wave numbers with a prime,
while obeying the constraint $\sum_j\bfk'_j+\sum_i\bfk_i=0$ (associated with
statistical homogeneity). 

When the soft wave numbers satisfy the hierarchy $k'_j \ll k'_{j+1}$ and
within the approximation $\Omega_{\rm}/f^2 \simeq 1$, the
angular-averaged consistency relation at equal times states that 
Eq.~(\ref{eq:angleaverage}) can be expressed in terms of the $n$th-order
polyspectrum as \cite{Valageas:2013lr,Kehagias:2013fk}

\be
\langle \overline{ \tdelta_{L,\bfk_1'} \dots \tdelta_{L,\bfk_{\ell}'} }
\tdelta_{\bfk_1} \dots \tdelta_{\bfk_n} \rangle'_{k'_j\to0} = 
\mathcal{L}'_1 \cdots \mathcal{L}'_\ell \cdot \; 
\langle \tdelta_{\bfk_1} \dots \tdelta_{\bfk_n} \rangle' . \nonumber \\
\label{eq:general}
\ee
In the right-hand side, the operators $\mathcal{L}_j'$ are given by
\be
\mathcal{L}'_j &=& P_L(k_j')\left[1 + \frac{13}{21}\frac{\partial}{\partial\ln D_+}\right.\nonumber\\
&&-\frac{1}{3}\sum_{m=j+1}^\ell\frac{\partial}{\partial\ln k'_m}
\left.-\frac{1}{3}\sum_{i=1}^n\frac{\partial}{\partial\ln k_i}\right],
\label{eq:operator}
\ee
where $P_L$ is the initial matter power spectrum linearly extrapolated to the time of
interest.
(Because these operators do not commute the ordering in the above relation 
only holds for the hierarchy of soft wave numbers $k'_1\ll k'_2\ll \dots\ll k'_\ell$.)
Because we take the limit $k_j'\to 0$ in Eq.(\ref{eq:general})
and we recover linear theory on large scales, we can replace the linear density
fields by the nonlinear ones and write
\be
\langle \overline{ \tdelta_{\bfk_1'} \dots \tdelta_{\bfk_{\ell}'} } \,
\tdelta_{\bfk_1} \dots \tdelta_{\bfk_n} \rangle'_{k'_j\to0} = 
\mathcal{L}'_1 \cdots \mathcal{L}'_\ell \cdot \; 
\langle \tdelta_{\bfk_1} \dots \tdelta_{\bfk_n} \rangle' . \nonumber \\
\label{eq:general-NL}
\ee

\subsection{Bispectrum} 
\label{subsec:bispec}
The simplest example of the relation~(\ref{eq:general-NL}) relates the angular-averaged
bispectrum to the power spectrum. This corresponds to $\ell=1$ and $n=2$, namely,
\be
\bar{B}(k';k)_{k'\rightarrow 0} & = & P_L(k')\left[1+\frac{13}{21}\frac{\partial}{\partial\ln D_+}-\frac{1}{3}\frac{\partial}{\partial\ln k}\right]P(k), \nonumber \\
&&
\label{eq:bispec}
\ee
where we denote
\be
\bar{B}(k';k)\equiv \langle \overline{\tdelta_{\bfk'}} \tdelta_{\bfk-\bfk'/2} 
\tdelta_{-\bfk-\bfk'/2} \rangle' , \hspace{0.3cm}
P(k) \equiv \langle \tdelta_{\bfk} \tdelta_{-\bfk} \rangle' ,
\label{eq:spec_def}
\ee
for the angular-averaged bispectrum and the power spectrum
[taking care of the constraint $\sum_j \bfk_j' + \sum_i \bfk_i = 0$ associated
with the Dirac factor in Eq.(\ref{eq:polyspec}) due to statistical homogeneity].
Because of statistical isotropy the spectra in Eq.(\ref{eq:bispec})
no longer have any dependence on the direction of $\bfk$.
Since higher-order polyspectra are increasingly noisy in general, in practice
the main application of these consistency relations is the lowest-order one 
(\ref{eq:bispec}), for the angular-averaged bispectrum.
We thus focus on the consistency relation~(\ref{eq:bispec}) in this study and
we test the low-$k'$ asymptotic behavior
with a large set of cosmological $N$-body simulations.

\subsection{Tree-level perturbation theory} 
\label{subsec:tree}
The relation~(\ref{eq:bispec}) has been checked by Ref.~\cite{Valageas:2013lr} at leading order of perturbation theory.
At this order, all we need is the second-order kernel of the matter density field
\cite{Bernardeau02}:
\be
F_2^s(\bfk_1,\bfk_2) &=& \frac{5}{7}
+ \frac{1}{2}\frac{\bfk_1\cdot\bfk_2}{k_1k_2}\left(\frac{k_1}{k_2}+\frac{k_2}{k_1}\right)
\nonumber\\
&&
+ \frac{2}{7}\frac{(\bfk_1\cdot\bfk_2)^2}{k_1^2k_2^2},
\label{eq:F2}
\ee
where we applied the approximation $\Omegam/f^2=1$. The time dependence of the kernel function~(\ref{eq:F2})
is actually very small \cite{Bernardeau02} and, for instance, approximately given 
by $(\Omegam^{-2/63}-1)$ 
in case of $\Omegam\simgt0.1$ for open universes without a
cosmological constant.
At tree order, the bispectrum, 
$B\equiv \langle \tdelta_{\bfk_1} \tdelta_{\bfk_2} \tdelta_{\bfk_3} \rangle'$, 
can be written as
\be
B(\bfk_1,\bfk_2,\bfk_3) = 2F_2^s(\bfk_1,\bfk_2)P_L(k_1)P_L(k_2) + \cyc,
\label{eq:Btree}
\ee
where $\cyc$ stands for two more terms given as the cyclic permutation over the three 
wave vectors. 
Then, taking the angular average of the tree-level bispectrum~(\ref{eq:Btree}) as
in Eq.(\ref{eq:spec_def}) gives
\be
\bar{B}(k';k) = P_L(k')\left[\frac{47}{21}-\frac{1}{3}\frac{\partial}{\partial\ln k}\right]P_L(k) + \mathcal{O}\left((k'/k)^2\right).
\label{eq:Bexpand}
\ee
Using $P_L(k,t) \propto D_+(t)^2$, this confirms the consistency 
relation~(\ref{eq:bispec}) within the validity of perturbation 
theory at the leading order.

\section{Simulation analysis}
\label{sec:sim}
We now describe the simulations that we analyze in this study. We also present
the method to measure the relevant statistical quantities and discuss the reliability
of the measurements. We finally show how accurately the consistency relation~(\ref{eq:bispec})
is recovered in the simulations.

\subsection{Setup of the simulations}
\label{subsec:nbody}
We use two sets of cosmological simulations in this paper. The first set of simulations
has already been used in Ref.~\cite{Taruya12}. 
Employing $1024^3$ particles, each of the $60$ independent
random realizations covers a comoving volume of $(2048\,h^{-1}\mathrm{Mpc})^3$. 
The total simulation volume of $515\,h^{-3}\mathrm{Gpc}^3$ enables 
precise measurements of statistical quantities.
These simulations are designed to calibrate analytical models of the matter power 
spectrum based on renormalized perturbation theory approaches on large scales 
(i.e., $k\simlt0.3\,\hmpci$) and the systematic error as well as the statistical error are 
controlled very well on these scales to meet the requirements (see also 
Refs.~\cite{Nishimichi09,Valageas11a} for more on the convergence).

However, because of their rather poor spatial resolution, it is known that 
the power spectrum on smaller scales is systematically smaller than it should be.
Although this systematic error is at most $\sim2\%$ at $k=0.4\,\hmpci$, almost
independently of redshift, it increases toward smaller scales.
The error reaches $4\%$ at $k\simeq0.7\,\hmpci$. 
Since our target accuracy in this study is about $5\%$ and, what is more, the consistency 
relation is less trivial on smaller scales (where we go beyond lowest-order perturbation
theory), we decided to run new simulations with a better spatial resolution.
We ran $512$ independent realizations of $512^3$-particle simulations, each of which 
had a cubic volume of $(512\,h^{-1}\mathrm{Mpc})^3$. 
This allowed us to double the dynamic range in wave number toward smaller scales, 
though the total simulation volume of these new simulations was only about $13\%$ 
of the low resolution simulations.
We examine in detail the convergence property of the spectra of interest in Appendix~\ref{app:conv}.
Based on the result, we adopt the simulations of Ref.~\cite{Taruya12} for the discussion on scales
$k\leq0.4\,\hmpci$, while the new high-resolution simulations are used on smaller scales.

\begin{figure} [htbp] 
   \centering
   \includegraphics[width=7cm]{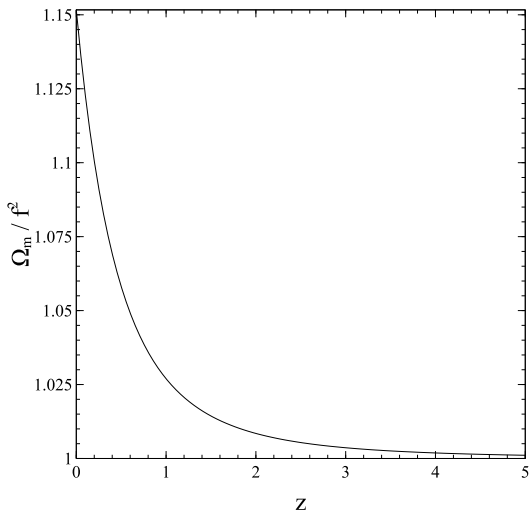} 
   \caption{Ratio $\Omegam/f^2$ for our cosmological model.}
   \label{fig:om_o_f2}
\end{figure}

The cosmological model in both set of simulations is a flat-$\Lambda$CDM model with the 
parameters $\Omegam=0.279$, $\Omegab/\Omegam=0.165$, $h=0.701$, 
$n_\mathrm{s}=0.96$, and $\sigma_8=0.816$, which is consistent with the five-year 
observation by the WMAP satellite \cite{WMAP5}. The combination $\Omegam/f^2$
in this cosmology is shown in Fig.~\ref{fig:om_o_f2}.
The ratio is very close to unity at high redshifts, $z \gtrsim 1$, and reaches
about $1.15$ at $z=0$.
In this paper, we test the consistency relation in our simulations at the
redshifts $z=1$ and $z=0.35$, at which the ratio $\Omegam/f^2$
departs from unity by $2.7\%$ and $7.5\%$, respectively.
However, the polyspectra at a given time are affected not only by the value of 
$\Omegam/f^2$ at that time but by its evolution history up to that epoch.
This further decreases the inaccuracy due to the approximation
$\Omegam/f^2 \simeq 1$ on the power spectrum and bispectrum, as found
in previous perturbative studies \cite{Bernardeau02}. We explicitly show the impact
of the breakdown of this approximation in a fully nonlinear manner in 
Sec.\ref{subsec:discussion},
by employing supplemental simulations done in the Einstein-de Sitter background.

\subsection{Left-hand side: Measurement of the bispectrum}
\label{subsec:LHS}
We first describe our method to measure the angular-averaged bispectrum in this subsection.
We assign particles onto $1024^3$ grid points using the cloud-in-cells (CIC) interpolation scheme 
(e.g., Ref.~\cite{Hockney81}) 
and apply the fast Fourier transformation to obtain the density field in Fourier space. 
We then correct the smoothing effect arising from the grid assignment by 
dividing by the CIC kernel function. We next take an average of cubic products of the density fields
to have an estimate of $\bar{B}$ defined in Eq.~(\ref{eq:spec_def}),
\be
\hat{B}(k';k) = \frac{V^2}{N_{k',k}^\mathrm{tri}}\sum_{\bfk'}\sum_{\bfk} \mathrm{Re}\left[\tdelta_{\bfk'} \tdelta_{\bfk-\bfk'/2} 
\tdelta_{-\bfk-\bfk'/2}\right],
\label{eq:Bestimate}
\ee
where $V$ stands for the simulation volume, $N_{k',k}^\mathrm{tri}$ is the number of triangles for the
wave number bin specified by $k'$ and $k$, and the summation is taken over modes $\bfk'$ and $\bfk$ that satisfy 
$k'-\Delta k'/2 \le |\bfk'| < k'+\Delta k'/2$ and $k-\Delta k/2 \le |\bfk| < k+\Delta k/2$, respectively.
We choose $\Delta k'= 0.004\,\hmpci$ and $\Delta k=0.02\,\hmpci$ for the low-resolution
simulations and $\Delta k'= 0.005\,\hmpci$ and $\Delta k=0.02\,\hmpci$ for the 
high-resolution ones.
Because we are now working on a periodic system with finite volume, the density field 
$\tdelta_{\bfk}$ is dimensionless, unlike the one in the previous section for continuous Fourier transforms.
In Eq.~(\ref{eq:Bestimate}), note also that we take the angular average not only over 
$\bfk'$ but also over $\bfk$,
in order to increase the statistics and suppress the statistical error level \footnote{
Once we take the angular average for $\bfk'$, $\bar{B}(k';k)$ no longer
has any angular dependence on $\bfk$ and thus the additional angular average over
$\bfk$ does not change the expectation value.}.

We finally take the average over different realizations to obtain our final estimate of the angular-averaged bispectrum
and we record the variance among realizations, divided by the square root of the number of realizations minus unity
(i.e., the standard error on the average values), to estimate the statistical error.

\begin{figure} [ht] 
   \centering
   \includegraphics[width=8cm]{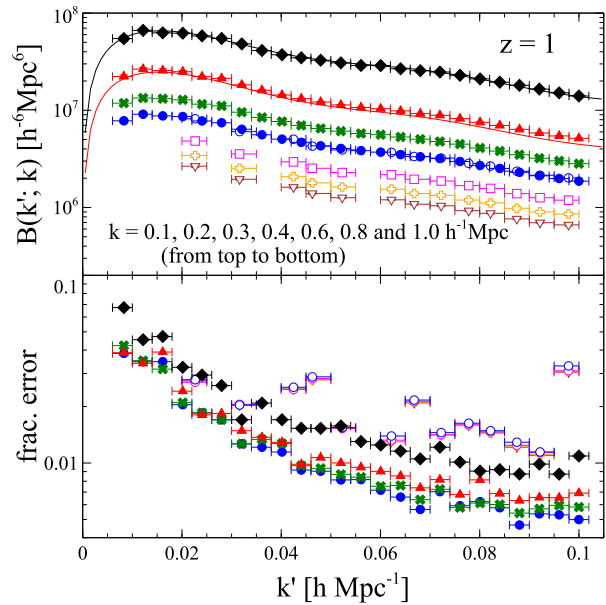} 
   \caption{Angular-averaged bispectrum at $z=1$. {\it Top:} measurements from $N$-body simulations
   (symbols) and the analytical predication at the tree level (lines; only for $k=0.1$ and $0.2\,\hmpci$). 
   The filled symbols depict the 
   measurements from lower-resolution simulations, while the open ones show those from 
   higher-resolution simulations. {\it Bottom:} fractional
   statistical error on the angular-averaged bispectrum estimated from the scatter among different random realizations.}
   \label{fig:LHS1}
\end{figure}
\begin{figure}[ht] 
   \centering
   \includegraphics[width=8cm]{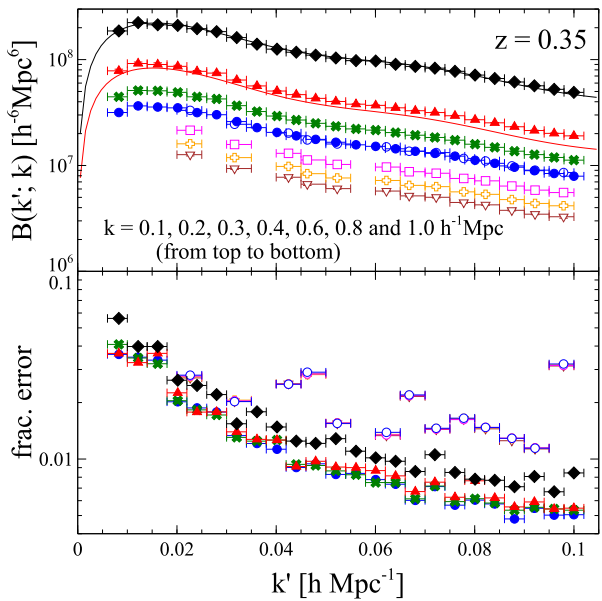} 
   \caption{Same as Fig.~\ref{fig:LHS1}, but at $z=0.35$.}
   \label{fig:LHS2}
\end{figure}

The resultant bispectrum is plotted in Figs.~\ref{fig:LHS1} and \ref{fig:LHS2} at $z=1$ and $0.35$, respectively.
We plot in the top panels the angular-averaged bispectrum, $\hat{B}(k'; k)$, as a function of wave number $k'$
for several fixed values of $k$ as written in the legend. The filled symbols show the measurements from
the low-resolution simulations, while the open ones depict those from the high-resolution simulations.

We also show with the solid line the perturbation-theory prediction at the tree level 
[i.e., Eq.~(\ref{eq:Btree})] for $k=0.1$ and
$0.2\,\hmpci$.
The measured angular-averaged bispectrum at $k=0.1\,\hmpci$ shows good agreement with perturbation theory, while the result at $k=0.2\,\hmpci$ reveals a lack of amplitude in the analytical curve.
This discrepancy is more important at $z=0.35$ ($10\%$ to $20\%$ depending on $k'$, and more evident at larger $k'$).
We omit analytical curves at $k\geq0.3\,\hmpci$ to avoid making the plot busy, but 
the discrepancy between the model and the simulations is even greater on these scales (a factor of $2$ or more).
Thus, we conclude that the applicable wave number range of the tree-level perturbation theory is limited to
$k\simlt 0.1\,h\,\mathrm{Mpc}$, both at $z=0.35$ and $1$.
In the top panels, we plot both filled and open circles at $k=0.4\,\hmpci$ to check the consistency 
between the two sets of simulations.
They are in agreement with each other from the comparison and further convergence tests are presented 
in Appendix~\ref{app:conv}.

Finally, the bottom panels of Figs.~\ref{fig:LHS1} and \ref{fig:LHS2} plot the fractional error on $\hat{B}(k'; k)$ measured
from the simulations (we adopt the same symbols as in the top panels).
Since we fix the bin width, $\Delta k'$ and $\Delta k$, the number of available Fourier-space triangles 
increases with $k'$ and $k$, resulting in a smaller error at smaller scale for filled symbols (i.e., low-resolution simulations). 
Also, the error level is higher for high-resolution simulations, which cover a smaller volume than the low-resolution ones.
The decrement of the error as a function of $k$ for the same set of simulations is only marginal, especially
at $z=0.35$, due to significant covariance among different modes on small scales. Eventually, at 
$k\simgt0.4\,\hmpci$, we do not observe clear dependence of the statistical error on $k$ for high-resolution simulations
(i.e., open symbols that are mostly overlapping with each other). On these scale, the statistical error is 
mostly determined by that in the soft mode $\tdelta_{\bfk'}$, and one does not gain much when one adds more 
hard modes $\tdelta_{\bfk}$.

The typical statistical error level on the angular-averaged bispectrum is roughly $1\%$, which allows us 
a meaningful test of the consistency relation. We are especially interested in $\bar{B}$ at the limit of small $k'$, and
the low-resolution simulations, which cover a total volume of $515\,h^{-3}\mathrm{Gpc}^3$, provide us
with measurements of the angular-averaged bispectrum down to $k'\sim 0.01\,\hmpci$ with
an error level of several percent. On the other hand, although the available data points are limited,
the high-resolution simulations enable us to test the consistency relation with a statistical error of $\sim3\%$ 
down to smaller scales where nonperturbative corrections to the density field
are important.

\subsection{Right-hand side: Measurement of the power spectrum and its derivatives}
\label{subsec:RHS}
We next describe our method to measure the right-hand side of Eq.~(\ref{eq:bispec}).
The three terms are explained one by one in the following, and we then summarize
the accuracy of the measurements of the sum of them.
\begin{figure} [ht] 
   \centering
   \includegraphics[width=8cm]{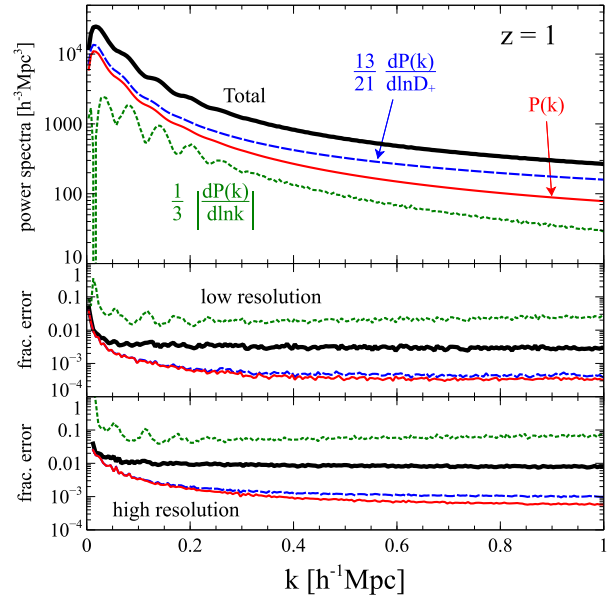} 
   \caption{Power spectrum and its derivatives at $z=1$. We plot the spectra from the low-resolution simulations
   in the top panel, and the statistical error is shown in the middle and the bottom panels, respectively, for the low-
   and high-resolution simulations.}
   \label{fig:RHS1}
\end{figure}
\begin{figure} [ht] 
   \centering
   \includegraphics[width=8cm]{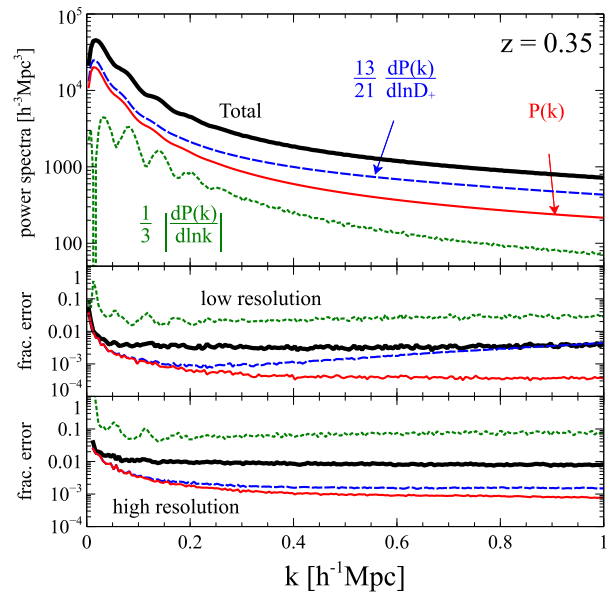} 
   \caption{Same as Fig.~\ref{fig:RHS1}, but at $z=0.35$.}
   \label{fig:RHS2}
\end{figure}

\subsubsection{Nonlinear power spectrum}
\label{subsubsec:RHS1}
The measurement of the nonlinear power spectrum is rather straightforward after
we have given the explanation for the bispectrum. The procedure is exactly the same as
in Sec.~\ref{subsec:LHS} up to the density field in Fourier space with the correction of the smoothing effect.
This time, we take
\be
P(k) = \frac{V}{N_{k}^\mathrm{mode}}\sum_{\bfk} \left|\tdelta_{\bfk}\right|^2,
\label{eq:Pestimate}
\ee
where $N_k^\mathrm{mode}$ stands for the number of Fourier modes in the $k$ bin.
In the summation, we consider modes $k-\Delta k/2 \leq |\bfk| < k+\Delta k/2$, and 
we adopt $\Delta k = 0.005\,\hmpci$ for both sets of simulations.
The results are shown by thin solid lines in the top panels of Figs.~\ref{fig:RHS1} and \ref{fig:RHS2}
at $z=1$ and $0.35$, respectively. We here plot the results of the low-resolution simulations,
but the high-resolution simulations almost coincide with the low-resolution simulations (see Appendix~\ref{app:conv}
for the convergence of the power spectrum). 

The statistical error on the measured power spectrum is plotted in the middle and the bottom panels for
the low-resolution and high-resolution simulations, respectively. Similarly to the bispectrum, the error level decreases
with wave number since we fix the bin width $\Delta k$ and thus we can access more Fourier modes at larger $k$.
Since the covariance between different modes grows with $k$ and time, the $k$ dependence of the fractional error is
shallower than $k^{-1}$ expected for uncorrelated measures.

\subsubsection{Time derivative}
\label{subsubsec:RHS2}
Estimating the time derivative of the power spectrum from the simulation data is less trivial.
We adopt the following procedure in this study.
Instead of preparing multiple snapshots at slightly different redshifts, we work on a
single snapshot of the positions and velocities of simulation particles.
We slightly displace the positions of particles according to their velocities:
\be
\bfx(t+\Delta t) = \bfx(a+\Delta a) = \bfx(t) + \mathcal{H}^{-1}(t)\bfv(t)\Delta a,
\label{eq:displace}
\ee
where $\bfx$ and $\bfv$ are the position and velocity of a particle in comoving coordinate and 
$\mathcal{H}=\rmd a/\rmd t$.
We repeat the same procedure as before and measure the power spectrum after applying the above 
displacement. We finally take the combination to estimate the derivative term:
\be
\frac{\rmd P(k)}{\rmd \ln D_+} = \frac{P(k; a+\Delta a/2)-P(k; a-\Delta a/2)}{\ln D_+(a+\Delta a/2)-\ln D_+(a-\Delta a/2)}.
\label{eq:timederiv}
\ee
This procedure can be justified as long as $\Delta a$ is small, and we adopt $\Delta a = 0.02$, which gives a converged result.

The measurement and its error are plotted in Figs.~\ref{fig:RHS1} and \ref{fig:RHS2} with the dashed line.
This term dominates the other terms over the whole range of wave numbers plotted in the figures.
The fractional error plotted in the bottom two panels behaves similarly to that on the nonlinear power spectrum at
small $k$ and is slightly larger on small scales reflecting the stronger nonlinearity in the momentum field than in the 
density field \footnote{Note that the time derivative of the density field is equivalent to the momentum field from the
continuity equation.}. 
\subsubsection{Wave number derivative}
\label{subsubsec:RHS3}
We compute the last term in the right-hand side of Eq.~(\ref{eq:bispec}) using the cubic spline fitting to the power 
spectrum measured above. Our choice of $\Delta k=0.005\,\hmpci$ is fine enough to evaluate 
the derivative without introducing a severe interpolation error. The measured derivative term shown with the dashed line
in Figs.~\ref{fig:RHS1} and \ref{fig:RHS2} exhibits a clear feature of baryon acoustic oscillations.
Note that we show the absolute value of this term as it is negative over most of the plotted wave number range.
The fractional error on this term estimated from the scatter among realizations is the largest among the three terms
probably because this term involves an interpolation and the derivative operation is not local in $k$, 
but the error level is still several percent over the most part of the plotted wave number range thanks to the large statistics.

\subsubsection{Sum of the three terms}
\label{subsubsec:RHS4}
Adding up the three terms already discussed and multiplying by the linear power spectrum, 
we finally obtain an estimate of the right-hand side of Eq.~(\ref{eq:bispec}).
We plot the sum of the three terms as the bold solid lines in Figs.~\ref{fig:RHS1} and \ref{fig:RHS2}.
The statistical error estimated from the scatter among realizations shown in the bottom two panels
is controlled below $1\%$ level both in the low- and high-resolution simulations on $k\simgt0.05\,\hmpci$.
This error level is in between that on the wave-number-derivative term (dotted) and the time-derivative term (dashed).
Since the former is the smallest among the three terms, its large error does not ruin the quality of the
sum of the three terms.
Thus the statistical error on the left-hand side (i.e., the angular-averaged bispectrum) of Eq.~(\ref{eq:bispec})
dominates over that in the right-hand side in checking the consistency relation in what follows.

In testing the relation~(\ref{eq:bispec}), we have to carefully take account of the effect of binning.
The left-hand side of the relation is binned both in $k$ and $k'$, while the right-hand side is written as 
a product of a $k$-binned quantity and the linear power spectrum $P_L(k')$.
Since the power spectrum has less statistical error than the bispectrum it can be measured in a 
finer binning as is done here. Then, we can basically interpolate the measured power spectrum 
and evaluate it at any $k$ without introducing a severe systematic error.
On the other hand, the linear power spectrum $P_L(k')$ could be obtained either from the definition of the cosmological model, without any measure from the simulations, or from the simulations. 
In Appendix~\ref{app:binning}, we explicitly show how this affects the comparison of the two sides, and we find that measuring all terms in the right-hand side from the simulations, with
a binning similar to the one used for the left-hand side, gives less noisy results
(because both sides now follow in the same manner the stochastic fluctuations of power 
from one realization to another).
Based on these results, we properly account for the binning both in $k$ and $k'$ for the right-hand side
to be consistent with that for the left-hand side.

\subsection{Results}
\label{subsec:results}
\begin{figure} [ht!] 
   \centering
   \includegraphics[width=8.3cm]{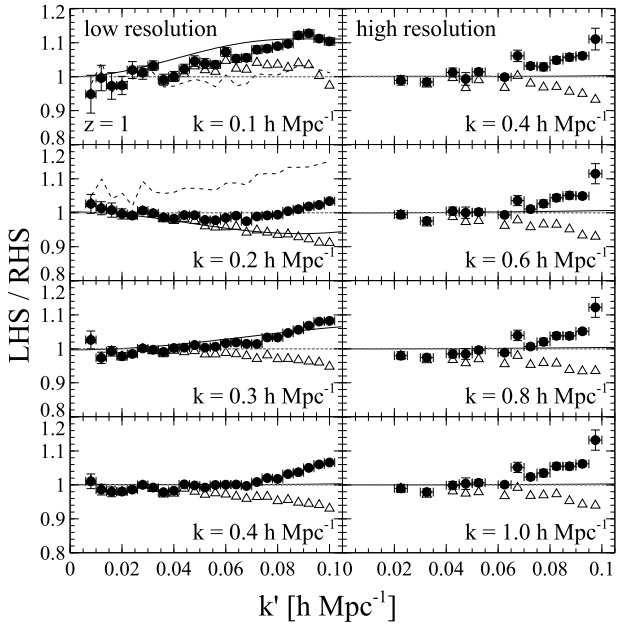} 
   \caption{Ratio of the two sides of Eq.~(\ref{eq:bispec}) at $z=1$. Each panel
   plots the ratio as a function of $k'$ for a fixed $k$ shown in the legend.
   The symbols are the results from low-resolution simulations (left panels)
   and high-resolution simulations (right panels).
   Filled circles correspond to measures of the bispectrum from 
   Eq.~(\ref{eq:Bestimate}), whereas for empty triangles we use for the soft mode the linear 
   density contrast $\tdelta_{L,\bfk'}$ instead of the nonlinear density contrast 
   $\tdelta_{\bfk'}$ as in Eq.(\ref{eq:general}). 
   The solid lines show the predictions of the tree-level perturbation theory 
   for this ratio, whereas the dashed lines in the left upper two panels show the
   ratio of the measured bispectrum to its tree-order prediction (\ref{eq:Btree}).} 
   \label{fig:ratio1}
\end{figure}

\begin{figure} [ht!] 
   \centering
   \includegraphics[width=8.3cm]{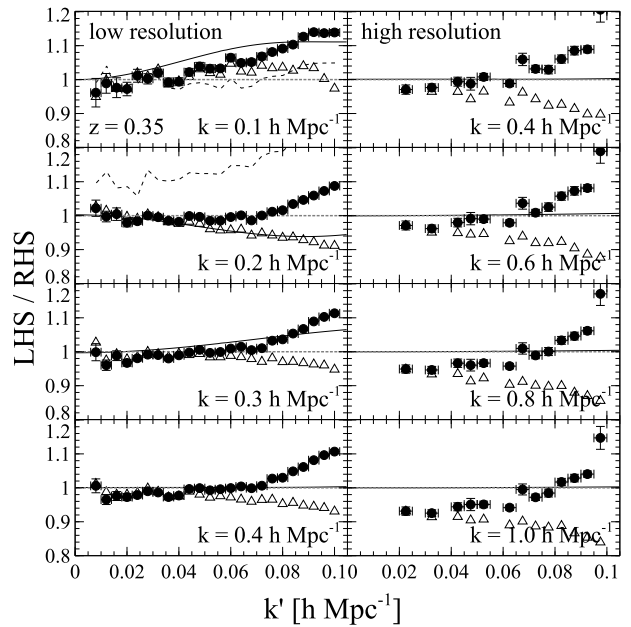} 
   \caption{Same as Fig.~\ref{fig:ratio1}, but the results at $z=0.35$.}
   \label{fig:ratio2}
\end{figure}

Now, we are in a position to discuss the validity of the consistency 
relation~(\ref{eq:bispec})
between the angular-averaged bispectrum and the power spectrum of the matter density field.
We consider the ratio of the two sides of Eq.~(\ref{eq:bispec}), measure this combination
from each realization, and then take the average over realizations, which is plotted in Figs.~\ref{fig:ratio1} and \ref{fig:ratio2},
respectively, at $z=1$ and $z=0.35$.


The left four panels in each of the two figures show the measurements from the low-resolution
simulations covering a larger volume ($0.1\,\hmpci\leq k \leq 0.4\,\hmpci$), 
while the right panels show those from high-resolution simulations 
($0.4\,\hmpci\leq k \leq 1.0\,\hmpci$), as a function of the soft wave number $k'$.
We also plot the ratio expected from the tree-level perturbation theory~(\ref{eq:Btree}) 
(solid lines) and the ratio of the measured bispectrum to the tree-order
prediction (\ref{eq:Btree}) in the left upper two panels (dashed lines).
The filled circles correspond to the bispectrum obtained from the nonlinear density 
fields measured at the redshift of interest, 
$\langle \overline{\tdelta_{\bfk'}} \tdelta_{\bfk-\bfk'/2} \tdelta_{-\bfk-\bfk'/2} \rangle'$
as in Eq.(\ref{eq:spec_def}), whereas the empty triangles correspond to the 
mixed bispectrum $\langle \overline{\tdelta_{L,\bfk'}} \tdelta_{\bfk-\bfk'/2} 
\tdelta_{-\bfk-\bfk'/2} \rangle'$, where we cross-correlate two nonlinear
fields with one linear field, as in Eq.(\ref{eq:general}).

In agreement with Figs.~\ref{fig:LHS1} and \ref{fig:LHS2}, the dashed lines
show that tree-level perturbation theory only gives an accurate prediction for the
bispectrum for $k'$ and $k$ below $\sim 0.1 \,\hmpci$.
When $k = 0.2 \,\hmpci$, it underestimates the bispectrum by about $10\%$,
and for higher $k$ the discrepancy becomes greater and can reach a factor of $2$
or more (it no longer appears in these panels because it is out of range).
This shows that the panels with $k \geq 0.3 \,\hmpci$ are beyond the
lowest-order perturbative regime and that we test the consistency relation
(\ref{eq:bispec}) in a nontrivial regime, beyond the perturbative check of
Sec.~\ref{subsec:tree}.

Even though lowest-order perturbation theory cannot predict the bispectrum
itself for $k  \gtrsim 0.2 \,\hmpci$, higher-order corrections partly cancel in the
ratio between both sides of Eq.~(\ref{eq:bispec}), and this ratio remains well
described by lowest-order perturbation theory up to $k'\sim 0.07\hmpci$ 
in all panels, where $k \leq 1 \,\hmpci$.
This also agrees with previous studies that found that the reduced bispectrum,
defined as $B(k_1,k_2,k_3)/[P(k_1) P(k_2) + {\rm (cyc.)}]$,
is more robust and shows smaller deviations from the perturbative
prediction than the bispectrum itself \cite{Bernardeau02}.
In particular, for $k \lesssim 0.3 \,\hmpci$, lowest-order perturbation theory
is able to reproduce the first deviations from unity of the consistency-relation
ratio, at $k' \sim 0.06 \,\hmpci$,
which may be either positive or negative, depending on scales.
In terms of the consistency relation (\ref{eq:bispec}), these departures signal
that the ratio $k'/k$ is not small enough to reach the low-$k'$ asymptotic behavior.
At higher $k'$, the behavior is the same in all panels, and the ratio grows with
$k'$. On the other hand, on large scales, $k' \lesssim 0.04 \,\hmpci$,
the ratio is consistent with unity.
We basically confirm the validity of the consistency relation~(\ref{eq:bispec}) within the statistical error
of the simulations, $\sim1$ to $5\%$, depending on the scales (see the leftmost data points in left panels).

Then, the results of the high-resolution simulations shown in the right panels, 
though we can sample a smaller number of data points, 
show a similar trend as that at $k=0.3$ or $0.4\,\hmpci$ found in the low-resolution simulations.
At the joint wave number, $k=0.4\,\hmpci$, the overall $k'$ dependence is consistent with the low-resolution ones:
the ratio is an increasing function of $k'$ and gradually deviates from unity at $k' \simgt 0.07\,\hmpci$.
At $z=1$, the $k'$ dependence in the other three panels is quite similar to that in the top right panel. 
The coherence of the zigzag pattern among the four panels might be explained by the fact that
we always use the same set of soft modes $\tdelta_{\bfk'}$ for different hard wave numbers.
However, we observe a systematic deviation from unity at small $k'$ at $z=0.35$.
The deviation is more prominent on larger $k$ and reaches up to $\sim7\%$ at $k=1\hmpci$ at that redshift.

Note that on these scales nonperturbative corrections such as shell crossing or 
the one-halo term in the halo model start to kick in (see, e.g., 
Refs.~\cite{Valageas11a,Valageas11b,Valageas13a,Valageas2013}).
However, in agreement with theoretical expectations, they do not lead to an
increasingly large deviation from unity of the low-$k'$ limit at $z=1$. Indeed, the
consistency relation (\ref{eq:bispec}) only relies on the approximate symmetry
associated with the approximation $\Omega_{\rm m}/f^2 \simeq 1$, and within this
approximation, it remains valid beyond shell crossing on highly nonlinear scales
for $k$. Nonlinearities might amplify the sensitivity to this approximation, but this
seems not to be the case in the range of scales shown in Figs.~\ref{fig:LHS1}. 
We will examine this issue more explicitly in the next subsection,
using additional simulations done in the Einstein-de Sitter
background at $z=0.35$.

The auto and mixed bispectra are consistent on large scales (i.e., small $k'$)
as we recover linear theory.
The differences that appear for $k' \gtrsim 0.06 \,\hmpci$ show that the soft mode 
density contrast $\tdelta_{\bfk'}$ begins to receive
non-negligible nonlinear corrections. These contributions violate the 
consistency relation because the latter is actually derived for the mixed polyspectra,
as in Eq.(\ref{eq:general}), and the form (\ref{eq:general-NL}) makes use of the
additional approximation $\tdelta_{\bfk'} \simeq \tdelta_{L,\bfk'}$.
Therefore, we would expect that the consistency relation is better satisfied
when we do not introduce this additional approximation and consider the mixed 
bispectrum, shown by the empty triangles.
The left panels do not show that the range of validity of the consistency relation is
extended when we use the mixed bispectrum, but this could be due to the fact that
the condition $k'\ll k$ is violated.
On the other hand, the right panels at $z=1$, with a lower scale ratio $k'/k$,
show a broader range of validity of the consistency relation when we use the mixed 
bispectrum, in agreement with these theoretical expectations.
The right panels at $z=0.35$ also show a broader plateau, as expected, but with
a small negative offset. Convergence studies presented in Appendix~\ref{app:conv} show
that this offset is not likely due to numerical error.
We thus conclude here that there exists a sign of a breakdown of the consistency relation~(\ref{eq:bispec})
on small scales at low redshift within the reliability of the present numerical simulations.
We will examine possible causes of the offset in the next subsection.

Note finally that we can see in some of the panels that filled circles are more consistent with unity
than empty triangles. However, this is just a coincidence: the downturn of the ratio is compensated
by the nonlinear correction to the soft mode $\tdelta_{\bfk'}$ by chance.
The mixed bispectrum is always a more direct measure of the consistency relation though it is not
an observable quantity. What we can do in practice is to push the measurement to the larger scale with larger surveys
to avoid nonlinear corrections to the soft mode.
 
\subsection{Effect of $\Omegam/f^2\neq1$}
\label{subsec:discussion}

One possible cause of the breakdown of the relation, which we find on small scales
at $z=0.35$, is the approximation $\Omegam/f^2 \simeq 1$ employed in deriving the consistency relations.
For instance, this combination deviates from unity by $\sim7.5\%$ at that redshift for the cosmological model we consider here,
as already discussed in Fig.~\ref{fig:om_o_f2}.
We conduct some supplemental simulations to understand to what extent this affects the spectra of interest.

\begin{figure} [ht!] 
   \centering
   \includegraphics[width=8.3cm]{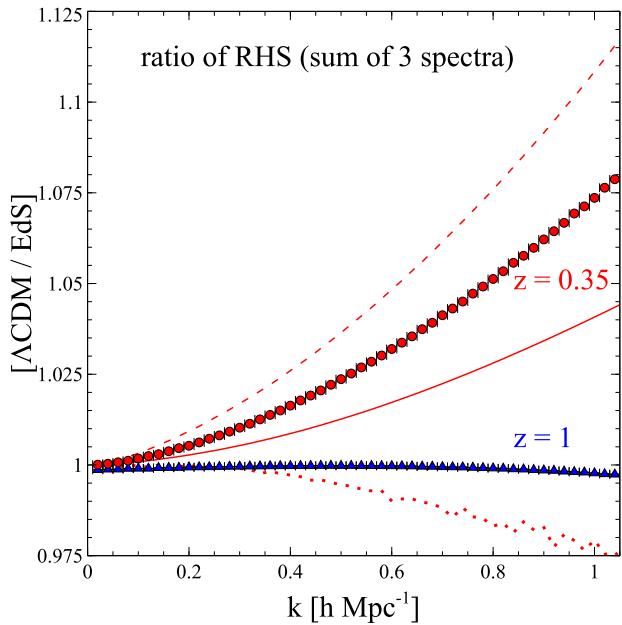} 
   \caption{Ratio of the right-hand side of the consistency relation~(\ref{eq:bispec}) in $\Lambda$CDM and EdS 
   background cosmology (triangles: $z=1$, circles: $z=0.35$). We also plot the contribution from each of the 
   three terms by lines at $z=0.35$: solid, dashed, and dotted lines, respectively, show the first, second, and the last terms.}
   \label{fig:EdS_pow}
\end{figure}

We run simulations with the same linear density field (i.e., the matter transfer function, 
the spectral tilt $n_\mathrm{s}$, and the amplitude of the linear fluctuations $\sigma_8$ at $z=0$) 
as the one we used in the main discussion, but adopting the cosmic expansion for the Einstein-de Sitter 
universe (i.e., $\Omegam=1$; EdS hereafter, where we also have $\Omegam/f^2=1$).
We simulate four random realizations of $1024^3$ particles in a cubic box with $2048\,h^{-1}\mathrm{Mpc}$ a side 
with the same random seeds as the first four realizations of the low-resolution simulations in the main discussion.

Although these supplemental simulations are not self-consistent in a sense that they still adopt 
the transfer function calculated for a $\Lambda$CDM model, they are useful to single out the 
effect of the breakdown of the approximation $\Omegam/f^2 \simeq 1$, by following the time evolution of the spectra
starting with exactly the same initial values but in different backgrounds that probe different ranges
of the ratio $\Omegam/f^2$.

Note that, even though the EdS universe satisfies $\Omegam/f^2=1$ at all times, at the background level, this is not sufficient to ensure the angular-averaged consistency relation is exact. Indeed, 
this relation assumes that $\Omegam/f^2 \simeq 1$ for all cosmologies close to the background one (as large-scale density fluctuations are identified with local changes of $\Omegam$).
Nevertheless, by changing the background cosmology, we change the reference point along the $\Omegam/f^2$ ratio as a function of density fluctuations, and the comparison between the $\Lambda$CDM and EdS cases gives an indirect probe of the effects associated with the approximation
$\Omegam/f^2 \simeq 1$.

Since the function $D_+(z)$ is different between the two models, we choose the
initial and output redshifts of the simulation in the EdS background such that they give the same 
linear growth rate: 
$D_{+,\mathrm{EdS}}(z_\mathrm{EdS}) = D_{+,\Lambda\mathrm{CDM}}(z_{\Lambda\mathrm{CDM}})$.
If the approximation is accurate enough, in the sense that the nonlinear power spectrum only depends on the linear growth rate, these simulations should give spectra 
indistinguishable to those from the simulations in the main discussion.
Also, we expect that the impact of any numerical error is almost the same in the two models
since we are simulating exactly the same stage of structure formation with the choice of redshifts above. 

We first show the ratio of the right-hand side of the relation~(\ref{eq:bispec}) in Fig.~\ref{fig:EdS_pow}.
We plot the ratio of the sum of the three terms by triangles and circles, respectively, at $z=1$ and $z=0.35$.
Although these symbols have vertical error bars estimated from the scatter among the four random realizations,
they are hardly visible by eye: they are typically $10^{-5}$ to $10^{-4}$ level. This ensures the robustness of
our estimate of the effect of $\Omegam/f^2\neq1$ using a rather small number of independent random realizations.

The plot shows that the effect is at most subpercent level at $z=1$, while a significant correction is observed at $z=0.35$.
The correction is an increasing function of $k$ and reaches to $\sim7.5\%$ at $k=1\,h\,\mathrm{Mpc}^{-1}$.
We also show the ratio for each of the three terms separately by lines 
(solid, dashed, and dotted for the first, second, and the last terms, respectively).
The solid line for the nonlinear matter power spectrum should be compared to similar analyses in the 
literature based on perturbation theories (e.g., Refs.~\cite{Takahashi:2008lr,Pietroni:2008qy,Hiramatsu:2009fk}).
Our simulation result is quantitatively in good agreement with these predictions in the literature on small scales
(i.e., $k\simlt 0.4\,h\,\mathrm{Mpc}^{-1}$). The wave-number-derivative term, depicted by the dotted line, is also 
affected by the non-EdS background at a similar level as the solid line but toward the opposite direction.
The largest effect lies in the time-derivative term (dashed).

\begin{figure} [ht!] 
   \centering
   \includegraphics[width=8.3cm]{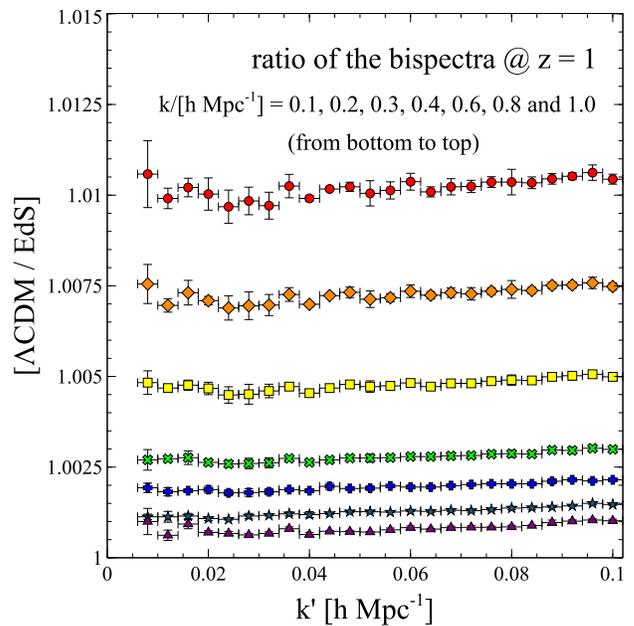} 
   \caption{Ratio of the angular-averaged bispectrum [left-hand side of the consistency relation~(\ref{eq:bispec})]
   in $\Lambda$CDM and EdS background cosmology at $z=1$.}
   \label{fig:EdS_bis_z1}
\end{figure}
\begin{figure} [ht!] 
   \centering
   \includegraphics[width=8.3cm]{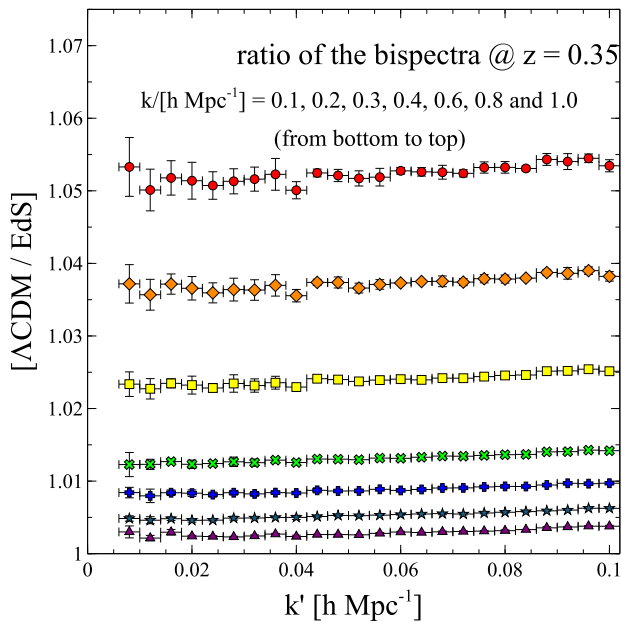} 
   \caption{Same as Fig.~\ref{fig:EdS_bis_z1}, but at $z=0.35$.}
   \label{fig:EdS_bis_z035}
\end{figure}

We now turn to the bispectrum and plot the ratio similar to the previous one in Figs.~\ref{fig:EdS_bis_z1} 
and \ref{fig:EdS_bis_z035}. We plot the ratio as a function of the soft wave number $k'$ for seven values of hard
wave number $k$ shown in the figure legend. Again, the ratio is always close to unity at $z=1$, and the deviation
is at most $\sim 1\%$ (at $k=1\,h\,\mathrm{Mpc}^{-1}$). On the other hand, the ratio can be as large as
$\sim1.05$ at $z=0.35$ on small scales. The size of the deviation from unity increases with the hard wave number $k$,
while its $k'$ dependence is weak at both redshifts.

An important observation here is that the correction from the non-EdS background
affects both sides of the consistency relation~(\ref{eq:bispec}), but its impact is quantitatively different in the two sides,
which do not cancel out when we take their ratio. 
At $k=1\,h\,\mathrm{Mpc}^{-1}$ at $z=0.35$, where we find that the consistency relation
holds the worst, the net effect to the consistency relation is to lower the ratio of the left- to right-hand side by about
$2\%$ to $3\%$ (see also Fig.~\ref{fig:EdS_ratio} below). 
This qualitatively explains the ratio of the two sides smaller than unity in Fig.~\ref{fig:ratio2}, 
though it does not completely recover the amplitude of the breakdown of the relation ($\sim 7\%$).
However, this is not surprising because the comparison between the $\Lambda$CDM 
and EdS cosmologies is only an indirect probe of the approximation $\Omegam/f^2\simeq 1$ 
(because neither of the two cosmologies is an exact reference point free from this approximation).
These results suggest that a percent-level breakdown of the relation is naturally expected at 
$z=0.35$ on small scales for the $\Lambda$CDM cosmology considered here at $z=0.35$.

\begin{figure} [ht!] 
   \centering
   \includegraphics[width=8.3cm]{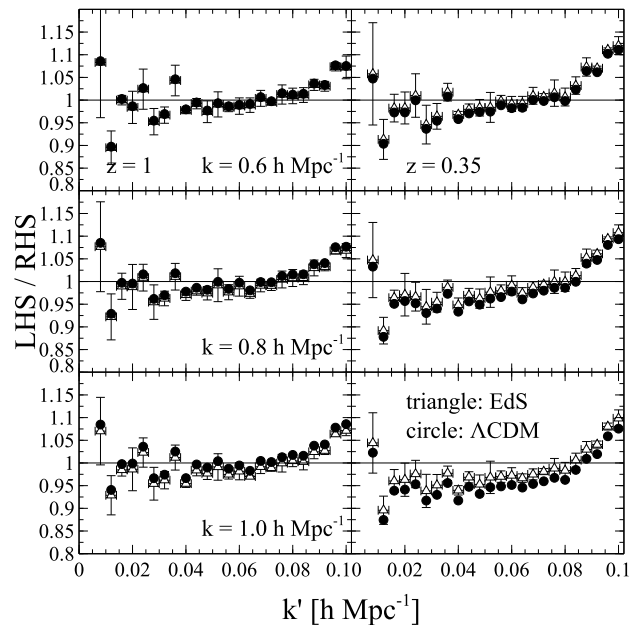} 
   \caption{Ratio of the two sides of the consistency relation~(\ref{eq:bispec}) in either 
   $\Lambda$CDM (circles) or EdS (triangles)
   background cosmologies, as in Figs.~\ref{fig:ratio1} and \ref{fig:ratio2}.
   We focus on large values of $k$ at $z=1$ (left column) and $z=0.35$ (right column).}
   \label{fig:EdS_ratio}
\end{figure}

We finally compare in Fig.~\ref{fig:EdS_ratio} the consistency-relation ratios obtained
in the $\Lambda$CDM and EdS cosmologies, focusing on the nonlinear scales
(we plot the data points from the four realizations for which the initial phases 
are reused in the EdS simulations to make the comparison fair).
We obtain qualitatively similar results for both cosmologies.
Moreover, in agreement with the previous figures, we find that for $k\simlt 1\,\hmpci$
at $z=1$ and $k\simlt 0.4\,\hmpci$ at $z=0.35$ the consistency relation is valid, whereas
for $\simgt 0.8\,\hmpci$ at $z=0.35$, the measured ratio is a few percent below unity, with
a slightly smaller departure for the EdS case.
As expected, the departure from unity takes place on scales where results from the EdS and 
$\Lambda$CDM cosmologies start to deviate.
To improve the theoretical predictions at low redshifts and small scales would require 
going beyond the approximation $\Omegam/f^2 \simeq 1$, but we do not investigate this problem here.

\section{Summary}
\label{sec:summary}
We have conducted a first numerical test of the angular-averaged consistency 
relation~(\ref{eq:general-NL}) 
by exploiting a large suite of cosmological $N$-body simulations. 
We focus on the lowest-order example of the relation ($\ell=1$ and $n=2$), 
which expresses the angular-averaged bispectrum in terms of the soft-mode and 
hard-mode power spectra.
The large total volume of the simulations allows us to conduct
a quantitative discussion on the validity of this relation.

We confirm that the relation is recovered within the statistical error of the simulations, 
beyond the validity range of the tree-level perturbation theory~(\ref{eq:Btree}), for 
$k\simlt 0.4\,\hmpci$ down to $z=0.35$. 
On the other hand, these scales remain within the range of higher-order perturbation theories so that 
the validity of the consistency relation is not surprising, because
it is well known that the approximation $\Omega_{\rm m}/f^2$ used in most
perturbative schemes is sufficiently accurate on these scales \cite{Bernardeau02}.
We indeed confirm that the breakdown of this approximation gives at most $\sim 1\%$ correction to
the spectra, and the effect on the power and bispectra mostly compensate with each other.

Beyond this regime, we find that the validity range of the consistency relation extends to 
smaller scales, up to $k \leq 1 \,\hmpci$ at $z=1$, where nonperturbative effects
are not negligible \cite{Valageas13a,Valageas2013}. 
We check that the condition $k'/k \ll 1$ is not sufficient for the consistency relation, and 
the soft mode $\tdelta_{\bfk'}$ must be in the linear regime.
Using the mixed bispectrum provides a more direct connection with the theory,
and our results suggest that this also extends the validity range of the consistency
relation. However, such a quantity can only be measured in numerical simulations
and not from observations of the real Universe.

We find, on the other hand, a statistically significant breakdown of the relation at $z=0.35$, where the angular-averaged
bispectrum is smaller than what we expect from the power spectrum.
This feature is more prominent on smaller scales reaching to $\sim7\%$ at $k=1\,\hmpci$.
We show that this is at least partly explained by the breakdown of the approximation 
$\Omegam/f^2\simeq 1$.
We present an extensive convergence study of simulations in Appendix~\ref{app:conv}, where we show
that the systematic error to the ratio of the two sides is only mild if any.
We thus conclude that the breakdown at this level should exist in reality in $\Lambda$CDM cosmologies 
with parameters consistent with recent observations at low redshifts.

We leave further discussions on, for instance, the effect of nonlinear bias or the 
usefulness of the relation to detect primordial non-Gaussianity to future studies. 
Also, it might be interesting to see how baryonic effects
alter the relation between different spectra in hydrodynamical simulations.

\acknowledgments
T.~N. is supported by Japan Society for the Promotion of Science (JSPS) Postdoctoral Fellowships
for Research Abroad. This work is supported in part by the French Agence Nationale de la Recherche under 
Grant No. ANR-12-BS05-0002.
The numerical calculations in this work were carried out on Cray XC30 at Center for 
Computational Astrophysics, CfCA, of National Astronomical Observatory of Japan.

\appendix
\section{Effect of binning}
\label{app:binning}
In practice, we have to adopt a binning in the wave number when we measure the power spectrum and the bispectrum.
Consequently, the consistency relation can only be tested between binned spectra.
In this Appendix, we briefly discuss the impact of binning and show the importance of a correct account of
this effect in testing the relations.

The angular-averaged bispectrum is naturally binned both in hard mode $k$ and soft mode $k'$.
On the other hand, the right-hand side of the relation consists of two separable factors, each of which
is a function of $k$ or $k'$. As for the $k$-dependent part, the statistical error level on the power spectrum and its
derivatives are smaller than those on the bispectrum 
even when we adopt a finer binning (compare Figs.~\ref{fig:LHS1} and \ref{fig:LHS2} with Figs.~\ref{fig:RHS1} 
and \ref{fig:RHS2}). This allows us to evaluate their values at an arbitrary wave number without introducing a severe
interpolation error. Moreover, the $k'$-dependent factor is simply the linear power spectrum 
$P_L(k')$,
which we do not need to measure since it is given from the beginning.
Thus, a simple way to evaluate the right-hand side is first to find the effective wave numbers $k$ and $k'$
at which the left-hand side is measured and then to evaluate the two factors composing the right-hand side at those
wave numbers.

\begin{figure} [!t] 
   \centering
   \includegraphics[width=7cm]{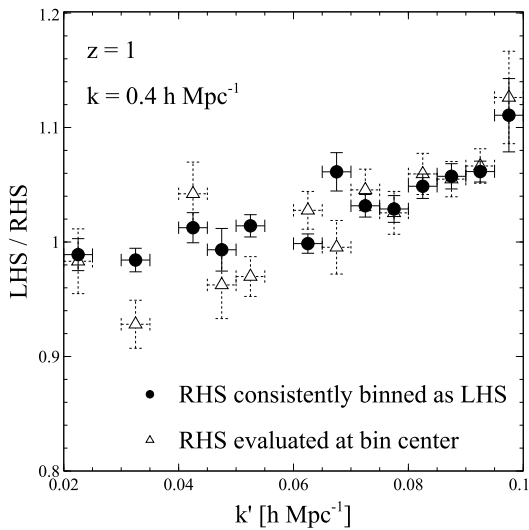} 
   \caption{Effect of binning on the test of consistency relations. 
   We plot the consistency-relation ratios measured by two different procedures: a)
   we take into account the binning in the right-hand side consistently to the left-hand side (circles), or b)
   we evaluate the right-hand side at the central wave number of each bin (triangles).}
   \label{fig:binning}
\end{figure}

The ratio of the two sides obtained this way is shown as open triangles with dashed error bars 
in Fig.~\ref{fig:binning} at $k=0.4\,h\,\mathrm{Mpc}^{-1}$ at $z=1$ from the higher-resolution simulations. 
In doing so, we adopt the mean over the norm of the wave vectors, $\bfk$ or $\bfk'$, that are taken into account 
for each bin and use the cubic spline interpolation
to evaluate the $k$-dependent factor at that representative $k$ value. 
The resultant data points exhibit rather noisy scatter around unity for which the
significance is larger than the statistical error level shown as error bars.

We next consider a binning scheme to the right-hand side, which is more consistent with the left-hand side.
For the $k$-dependent factor, we simply apply the same bin width as in the bispectrum measurement and
measure the binned nonlinear power spectrum and its derivatives.
In addition, we now measure the linear power spectrum from the random linear density fields used in setting
up the initial conditions of the simulations in consistent $k'$ bins
instead of evaluating the true spectrum defined as an ideal ensemble average.
Since the summation in Eq.~(\ref{eq:Bestimate}) over $\bfk'$ is taken only for $2k_\mathrm{f}$ times integer
vectors, where $k_\mathrm{f}=2\pi/L_\mathrm{box}$ is the fundamental wave number, such that $\bfk'/2$ is available 
in the simulations, we consider only these wave numbers when we measure the binned linear power spectrum.

The resultant ratio of the two sides is plotted as filled circles in Fig.~\ref{fig:binning}.
Now, the data points look cleaner than the triangles obtained with unbinned right-hand side.
Also, the error bars on circles are significantly smaller than those on triangles in most of the cases.
This is because the randomness in the angular-averaged bispectrum arising from the finite simulation box
is partly cancelled by the linear power spectrum which now takes account of 
the same randomness in each of the realizations.
From these considerations, we adopt the latter binning scheme for the right-hand side in the main text.

\section{Convergence study}
\label{app:conv}
\begin{table}[!t]
   \centering
   \caption{List of simulations for the convergence study. The box sizes of the simulations are in units of $\himpc$. Note
   that the simulations subhigh are a subset of main high.}
   \begin{tabular}{l||cccc}
   Name & Box size & Particles & $z_\mathrm{in}$ & Realizations \\
   \hline
   Main low & 2048 & $1024^3$ & 15 & 60 \\
   Main high & 512 & $512^3$ & 31 & 512 \\
   Sublow & 512 & $256^3$ & 15 & 4 \\
   Subhigh & 512 & $512^3$ & 31 & 4 \\
   Ref. & 512 & $1024^3$ & 63 & 4
   \end{tabular}
   \label{tab:conv}
\end{table}

We study the convergence properties of the power and bispectrum measured from $N$-body simulations in this Appendix.
For this purpose, we have run extra simulations with different spatial resolutions.
We list the parameters for these simulations (``sublow", ``subhigh" and ``ref.") in addition to 
the two sets of simulations presented in the main text 
(``main low" and ``main high") in Table~\ref{tab:conv}.
All the three sets of supplemental simulations have the same volume ($512^3\,h^{-3}\mathrm{Mpc}^3$) but
have different spatial/mass resolutions.
We use them to understand the systematic error caused by the finite resolution.

We also adopt different starting redshifts of the simulations for these three.
They are determined to minimize the sum of the two systematic errors: 
the transient effect caused by the initial conditions created with the Lagrangian perturbation theory and the inaccuracy of 
the tree force in the early stages of the simulations where particles are distributed closely to a regular grid \cite{Taruya12}.
In other words, we can safely start the simulations at a high redshift only when the spatial resolution is sufficient to
control the force accuracy.
Since the relative displacement of particles with respect to the grid spacing depends on the spatial resolution, the
optimal redshifts vary with resolutions.
Thus, we are testing the combination of the transient effect and the resolution effect by comparing different sets of simulations.

Note that the two simulations, sublow and subhigh, respectively, have the same resolutions as 
main low and main high simulations presented in the main text. Indeed, sub high is a subset of four realization
from the main high simulations. We are thus testing the systematic error in the main two sets of simulations using
the supplemental simulations and comparing with the reference simulations with resolution twice better than the 
higher-resolution simulations.

We adopt exactly the same initial random phases for the three sets of supplemental simulations, and we conduct
four realizations for each of them to estimate the statistical scatter. We will show shortly that we can indeed achieve a
small statistical error on the ratio of the same spectrum from different simulation setups with a rather small number
of independent realizations. 

\begin{figure} [!t] 
   \centering
   \includegraphics[width=7cm]{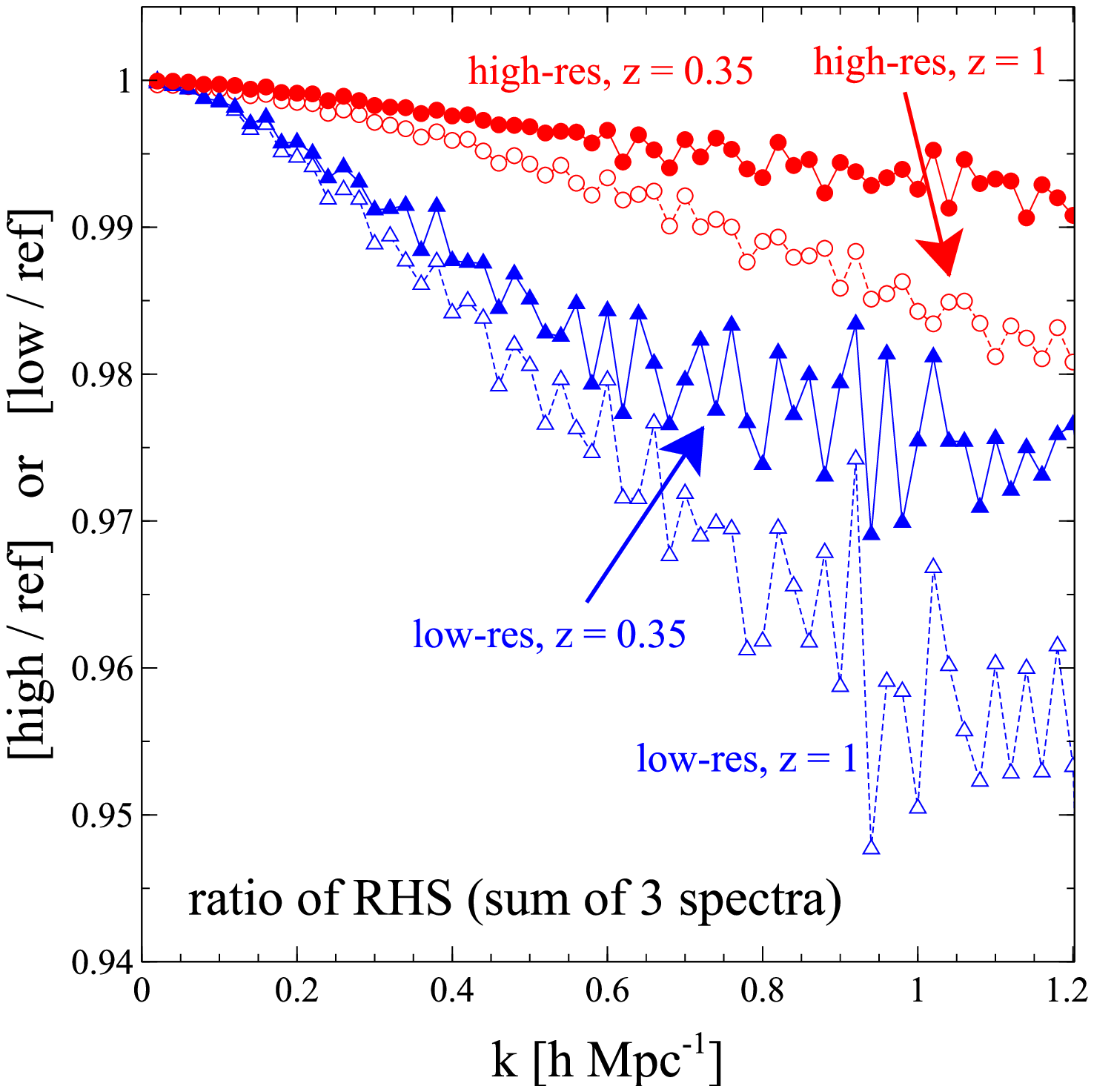} 
   \caption{Convergence of the right-hand side of the consistency relation~(\ref{eq:bispec}) 
   against resolution and the starting redshift of simulations. We show the sublow (subhigh) 
   simulations divided by ref. simulations in triangles (circles). Filled symbols are the ratio at $z=0.35$,
   while we plot the results at $z=1$ by open symbols.}
   \label{fig:convP}
\end{figure}

We first show in Fig.~\ref{fig:convP} the right-hand side of the relation~(\ref{eq:bispec}), 
the sum of the power spectrum and its derivative terms measured from sub low and sub high simulations, 
divided by that from the reference simulations that have the best spatial resolution. 
Shown by triangles are the results of the low-resolution runs, while circles are the results of the high-resolution runs.
Filled (open) symbols depict data from the outputs at $z=0.35$ ($1$).

We can learn from the figure that the systematic error grows with wave number and decreases with time.
This indicates that the effect is transient and likely comes from the inaccuracy in the initial condition set by 
second-order Lagrangian perturbation theory (2LPT).
The higher-resolution simulations, which also start at a higher redshift, are less affected by this systematics than
the lower-resolution ones. The error gets larger and reaches to $\sim 2\%$ and $1\%$ at $k=1\hmpci$ for the 
higher-resolution simulations, while the lower-resolution simulations can have a $\sim5 \%$ error in the worst case.

\begin{figure} [!h] 
   \centering
   \includegraphics[width=7cm]{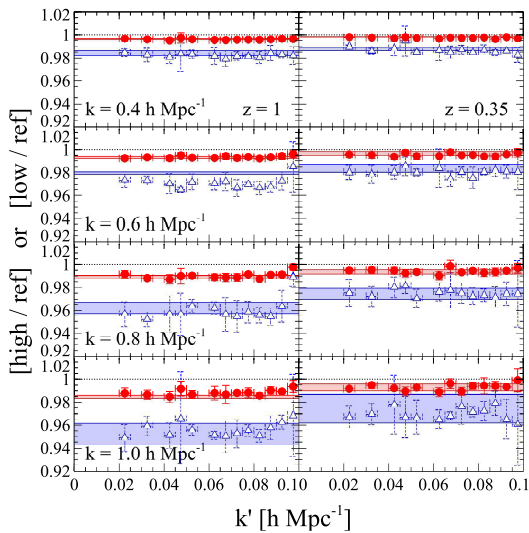} 
   \caption{Convergence of the left-hand side of the consistency relation~(\ref{eq:bispec})
   against resolution and the starting redshift of simulations.
   We show the sub low (sub high) simulations divided by ref. simulations in triangles (circles).
   Also plotted in horizontal bands are the same ratio but for the right-hand side at the wave number $k$
   shown in the figure legend (see Fig.~\ref{fig:convP}).
}
   \label{fig:convB}
\end{figure}

We then present the ratio between the same set of simulations but of 
the angular-averaged bispectrum in Fig.~\ref{fig:convB} at $z=1$ (left)
and $z=0.35$ (right). Again, the circles and triangles, respectively, show the result of the high- and low-resolution
simulations divided by the measurement from the reference simulations.
Also shown by horizontal bands are the ratio of the right-hand side at the wave number $k$ indicated in the figure legend, 
which we have just discussed in Fig.~\ref{fig:convP}. Interestingly, the systematic effect on the bispectrum (symbol)
is at a similar level as that on the combination of power spectrum (bands).
Thus, this effect is not likely to change significantly the relation between the two sides of Eq.~(\ref{eq:bispec}), and
the possible net effect on the relation is at most $1\%$ ($k=0.6\hmpci$ at $z=1$ for the lower-resolution simulations).
To be conservative, however, we show only the higher-resolution simulations on smaller scales ($k\geq0.6\hmpci$)
in the main text. 

\begin{figure} [!t] 
   \centering
   \includegraphics[width=7cm]{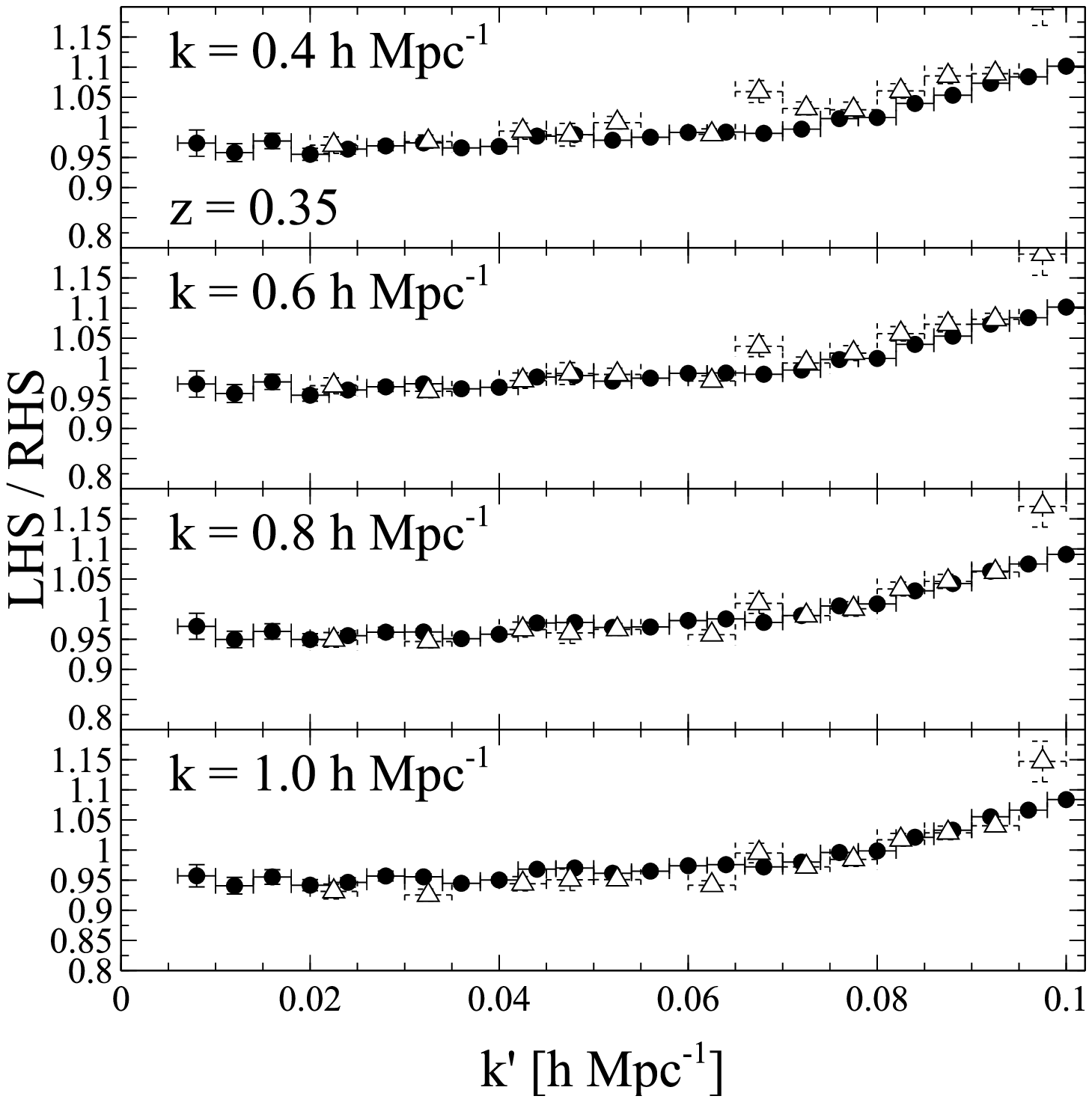} 
   \caption{Convergence of the consistency-relation ratio against the simulation volume at $z=0.35$. 
   We plot ``main low" (``main high") by circles (triangles).}
   \label{fig:conv_vol}
\end{figure}

The results so far indicate that the finite spatial resolution as well as the transient effect caused by the 2LPT initial condition
are not responsible for the breakdown of the consistency relation~(\ref{eq:bispec}) seen in Fig.~\ref{fig:ratio2} at $z=0.35$. 
Given that, the comparison between two simulations in different volumes and with different resolutions allows us to 
test the effect of the finite simulation volume, since the impact of the latter is shown to be rather small.  
We plot the ratio of the two sides of the relation~(\ref{eq:bispec}) at $z=0.35$ in 
Fig.~\ref{fig:conv_vol} measured from the two sets of simulations used in the main discussion.
The two symbols (circles: main low and triangles: main high) mostly lie close to each other.
The difference between the two symbols is typically a few percent level, which is similar to the size of the statistical error.
This shows that the finiteness of the simulation volume does not severely affect the consistency relation~(\ref{eq:bispec}).

\bibliography{LSSref}

\end{document}